\documentclass{article}

\usepackage{arxiv}

\usepackage[utf8]{inputenc} 
\usepackage[T1]{fontenc}    
\usepackage{hyperref}       
\usepackage{url}            
\usepackage{booktabs}       
\usepackage{amsfonts}       
\usepackage{nicefrac}       
\usepackage{microtype}      
\usepackage{lipsum}
\usepackage{graphicx}
\graphicspath{ {./images/} }

\usepackage[numbers]{natbib}
\usepackage{amsmath}
\usepackage{caption}
\usepackage{subcaption}
\usepackage{url}

\hypersetup{
	colorlinks=false,
	pdfborder={0 0 0},
	pdfborderstyle={/S/U/W 0},
}

\title{Sequential Permutation Testing of Random Forest Variable Importance Measures}

\author{
 Alexander Hapfelmeier \\
  Institute of General Practice and Health Services Research \\
  Institute of AI and Informatics in Medicine \\
  Technical University of Munich \\
  \texttt{Alexander.Hapfelmeier@tum.de} \\
   \And
 Roman Hornung \\
  Institute for Medical Information Processing, Biometry and Epidemiology \\
  University of Munich \\
  \texttt{hornung@ibe.med.uni-muenchen.de} \\
  \And
 Bernhard Haller \\
  Institute of AI and Informatics in Medicine \\
  Technical University of Munich \\
  \texttt{Bernhard.Haller@tum.de} \\
}

\begin{document}
\maketitle
\begin{abstract}
Hypothesis testing of random forest (RF) variable importance measures (VIMP) remains the subject of ongoing research. Among recent developments, heuristic approaches to parametric testing have been proposed whose distributional assumptions are based on empirical evidence. Other formal tests under regularity conditions were derived analytically. However, these approaches can be computationally expensive or even practically infeasible. This problem also occurs with non-parametric permutation tests, which are, however, distribution-free and can generically be applied to any type of RF and VIMP. Embracing this advantage, it is proposed here to use sequential permutation tests and sequential p-value estimation to reduce the high computational costs associated with conventional permutation tests. The popular and widely used permutation VIMP serves as a practical and relevant application example. The results of simulation studies confirm that the theoretical properties of the sequential tests apply, that is, the type-I error probability is controlled at a nominal level and a high power is maintained with considerably fewer permutations needed in comparison to conventional permutation testing. The numerical stability of the methods is investigated in two additional application studies. In summary, theoretically sound sequential permutation testing of VIMP is possible at greatly reduced computational costs. Recommendations for application are given. A respective implementation is provided through the accompanying R package $rfvimptest$. The approach can also be easily applied to any kind of prediction model. 
\end{abstract}


\section{Introduction}
\label{sec:intro}
Random forests (RF) \citep[]{Breiman2001} are used in various fields to solve regression, classification and time-to-event problems because the machine learning method has many appealing properties. It has been shown to rank among the best performing prediction models \citep[]{Fernandez2014, Wainberg2016}, while offering high flexibility due to its applicability to predictor variables and outcomes of any scale, to potentially non-linear and interacting relationships between variables \citep[]{Strobl2009, Boulesteix2015}, in the presence of missing data \citep[]{Hapfelmeier2012}, and in high-dimensional data settings \citep[]{Boulesteix2012, Belgiu2016}. The algorithm has also been the subject of continuous development since its introduction. Thus, it has been extended for quantile regression \citep[]{Meinshausen2006}, ordinal regression \citep[]{Hornung2020, Tutz2021} and model-based recursive partitioning \citep[]{Garge2013, Seibold2018, Hothorn2021}, among many others.

Another important feature of the algorithm is the variable importance measures (VIMP), which are designed to quantify the relevance of a predictor variable to the prediction problem. The Gini VIMP and the permutation VIMP were originally proposed and have been the most commonly used measures since then \citep[]{Breiman2001, Strobl2009, Boulesteix2012}. However, both measures were found to be biased and corrections were suggested \citep[]{Hothorn2006, Strobl2007, Strobl2007b}. Alternate ways of measuring variable importance have also been proposed, where variables are removed, replaced (e.g. with knockoff variables) or conditionally permuted in the data used for prediction by a given RF or used for building a compared RF \citep[cf.][for a more detailed discussion]{Hooker2021}. In addition to pure quantification and the resulting ranking regarding the relevance of a predictor variable in the prediction model, VIMP were also used for variable selection \citep[]{Hapfelmeier2013, Hapfelmeier2014, Degenhardt2019, Speiser2019, Bommert2020}. 

Some of the proposed variable selection methods are filtering methods based on hypothesis testing. Such testing can generically be implemented by permutation tests for any type of RF and VIMP. However, permutation testing entails the problem of efficiency \citep[]{Hapfelmeier2013, Nembrini2018, Speiser2019}. Especially in high-dimensional data settings the computational burden can become very high. Alternative proposals that base on bootstrapping, subsampling and jackknife estimation \citep[]{Ishwaran2019} or analytically derived formal hypothesis testing of variable importance under regularity conditions \citep[]{Mentch2016, McAlexander2020} share the same problem to some extent. Other sophisticated heuristic approaches make assumptions on the distribution of VIMP which are mainly based on empirical evidence \citep[]{Janitza2018} and lack formal proof \citep[]{Ishwaran2019}. 

In the present work, we propose the use of sequential permutation tests and p-value estimation to reduce the high computational cost of permutation tests for VIMP \citep[][]{Wald1945, Lock1991, Besag1991}. Thereby, a hypothesis can be accepted or a p-value can be estimated prematurely, that is before all permutations of a permutation test have been processed. The basic idea of the sequential permutation test is that it can become clear early on which hypothesis is likely to be accepted in the end. The tests have been designed to control the type-I error probability at a nominal level while preserving or controlling a high power, in terms of a low type-II error probability, at a remarkably reduced computational cost. These theoretical properties are presented and discussed in the context of hypothesis testing of the popular and widely used permutation VIMP. Their consistency with the results of simulation studies and applications to real data, as well as the variability of the non-deterministic results of the permutation tests, which is due to the randomness involved in the computations, are investigated. Recommendations for application of the different sequential procedures are derived from a discussion of these findings.

\section{Methods}
\label{sec:meth}

\subsection{Sequential permutation testing}
\label{sec:permtest}

Permutation tests first emerged in the 1920s and were initially used to validate the results of parametric tests \citep[]{Berry2011}. They were suitable for this purpose because they are not only accurate and unbiased, but also do not require any distributional assumptions and provide exact p-values \citep[]{Lehmann2005, Good2013}. The general principle of a one-sided permutation test of a null hypothesis against a simple alternative is to compute an appropriate statistic $u$ from the original data. The rank of $u$ is then determined in the set of similar statistics $u_{s}$, $s \in \{1, \ldots, n!\}$, computed using all possible arrangements or permutations of the $n$ observations of the original data. For example, a one-sided null hypothesis could be rejected if 
\begin{equation} \label{eq:pval_orig}
\frac{\sum_{s = 1}^{n!} I(u_{s} \geq u)}{n!} \leq \alpha,
\end{equation}
where the indicator function $I()$ takes the value $1$ whenever the condition in parantheses is true and $0$ otherwise. However, an apparent disadvantage of permutation tests is their high computational cost. Given $n$ observations, there are $n!$ possible arrangements of the data, which require a corresponding number of permutations and calculations of the statistics $u_{s}$ to be done. 

Therefore a size-$\alpha$ test has been suggested which is based on a Monte Carlo random set of all possible permutations of size $M<<n!$. It was shown that the power of such a modified test decreases at most by a factor of 0.922, 0.945 and 0.980 for $M=500$, $M=1000$ and $M=10000$ when $\alpha=0.05$ \citep[]{Dwass1957}. The power of this modified test decreases further with more conservative values of $\alpha<0.05$. 

\subsubsection{Certain premature termination}

With a fixed maximum number of $M$ permutations, this test procedure can even be terminated prematurely at a stage $m<M$. It is therefore obvious that $H_{0}$ cannot eventually be rejected if a sufficient number of statistics $u_{s}$ greater than or equal to the original statistic $u$ have already been observed within the first $m$ permutations \citep[]{Wald1945, Good2006}, that is if 
\begin{equation} \label{eq:earlystopp1}
\frac{\sum_{s=1}^{m}I(u_{s} \geq u)}{M} > \alpha.
\end{equation}
Likewise, $H_{1}$ can be accepted prematurely if a significant result cannot become insignificant even if the remaining $M-m$ statistics $u_{s}$ were greater than or equal to the original statistic $u$, that is if 
\begin{equation} \label{eq:earlystopp2}
\frac{\sum_{s=1}^{m}I(u_{s} \geq u) + (M-m)}{M} \leq \alpha.
\end{equation}
In addition to these obvious stopping rules (\ref{eq:earlystopp1}) and (\ref{eq:earlystopp2}), more sophisticated approaches have been proposed that minimise the expected number of permutations required until the process terminates prematurely, while controlling the type-I and type-II error probabilities.

\subsubsection{Sequential probability ratio test}

In a seminal paper on sequential tests of statistical hypotheses, \cite{Wald1945} introduced the sequential probability ratio test (SPRT) for permutation testing. When testing a hypothesis $H_{0}$ against a single alternative $H_{1}$, it is declared to be optimal or most efficient compared to other sequential hypothesis tests, or the modified and current most powerful test which uses a fixed number of $M$ permutations, because it requires the smallest expected number of permutations while controlling for type-I and type-II error probabilities equally. Considering a random variable $Z$ with probability density function $f(z,\theta)$ and the unknown parameter $\theta$, the SPRT can be used for sequentially testing $H_{0}{:}\; \theta = \theta_{0}$ against the alternative $H_{1}{:}\; \theta = \theta_{1}$ with $\theta_{1} < \theta_{0}$.

Consider a binomial distribution $f(z,m,\theta)=f(z,m,p)=\binom{m}{z}p^{z}(1-p)^{m-z}$, $z \in \{1,2,\ldots,m\}$, of a random variable $Z$. In the present setting, $z = \sum_{s=1}^{m} I(u_{s} \geq u)$ is a realization of $Z$, often referred to as $d_{m}$. It simply specifies the number of statistics $u_{s}$ that are greater than or equal to the original statistic $u$ at stage $m<M$ of a permutation test. A SPRT of $H_{0}{:}\;p=p_{0}$ against $H_{1}{:}\; p=p_{1}$ with $0<p_{1}<p_{0}<1$ and controlled type-I and type-II error probabilities $\alpha$ and $\beta$ suffices to test $H_{0}{:}\;p=p_{0}$ against $H_{1}{:}\; p<p_{0}$ in such a case as the type-II error probability is a monotonically decreasing function of $p$ when $p < p_{0}$. It must therefore also apply that the type-I error probability is less than or equal to $\alpha$ and the type-II error probability is less than or equal to $\beta$ when $p \geq p_{0}$ or $p \leq p_{1}$, respectively, for such a test. 

The SPRT is carried out by accepting $H_{0}$ at stage $m <M$ when
\begin{equation} \label{eq:reject}
\frac{p_{1m}}{p_{0m}} \leq \frac{\beta}{1-\alpha}
\end{equation} 
or by accepting $H_{1}$ at stage $m <M$ when
\begin{equation} \label{eq:accept}
\frac{p_{1m}}{p_{0m}} \geq \frac{1-\beta}{\alpha},
\end{equation} where
\[  
\frac{p_{1m}}{p_{0m}} = \frac{p_{1}^{d_{m}}(1-p_{1})^{m-d_{m}}}{p_{0}^{d_{m}}(1-p_{0})^{m-d_{m}}}.
\] 
When no hypothesis can be accepted at this stage another permutation is performed and the testing procedure is repeated. Solving the inequalities (\ref{eq:reject}) and (\ref{eq:accept}) by $d_{m}$ leads to the respective decision rules to accept $H_{0}$ at stage $m<M$ when
\begin{equation} \label{eq:acceptsuccesses}
d_{m} \geq \frac{log(A)+m\ log\!\left(\frac{1-p_{0}}{1-p_{1}}\right)}{log\!\left(\frac{p_{1}(1-p_{0})}{p_{0}(1-p_{1})}\right)}
\end{equation} 
and to accept $H_{1}$ at stage $m<M$ when
\begin{equation} \label{eq:rejectsuccesses}
d_{m} \leq \frac{log(B)+m\ log\!\left(\frac{1-p_{0}}{1-p_{1}}\right)}{log\!\left(\frac{p_{1}(1-p_{0})}{p_{0}(1-p_{1})}\right)},
\end{equation}
where $A = \frac{\beta}{1-\alpha}$ and $B = \frac{1-\beta}{\alpha}$. The parameters determining the decision rules are therefore $(p_{0}, p_{1}, \alpha, \beta)^{\top}$ or equivalently $(p_{0}, p_{1}, A, B)^{\top}$.

\subsubsection{Sequential method for approximating a general permutation test}

Building on the SPRT, \cite{Lock1991} suggested a sequential method for approximating a general permutation test (SAPT). It was designed to strictly control the type-I error probability at an effective significance level close to $\alpha$ while achieving a power close to that of the exact permutation test and requiring only a minimal number of permutations. Key to the development of SAPT was the estimate of the effective significance level with appropriate bounds derived from the power function of the SPRT. \cite{Lock1991} infers that choosing values for $p_{0}$ and $p_{1}$ that are equidistant from a nominal significance level $\alpha$ such that $(p_{0} + p_{1})/2 = \alpha$, and a small $A$ with the condition that $B=1/A$, leads to a test with a conservative effective significance level close to $\alpha$. With smaller differences between $p_{0}$ and $p_{1}$ and smaller values of $A$, SAPT requires more permutations but also gains power. In a range of investigated values $p_{0} = 0.06$, $p_{1} = 0.04$ and $A = 0.1$ seemed to be reasonable choices leading to an effective significance level close to $0.0498$ when aiming for a nominal significance level of $\alpha = 0.05$. Accordingly, values of $p_{0}$ and $p_{1}$ more distant from 0.05 resulted in an even lower effective significance level.

\begin{figure}
     \centering
     \begin{subfigure}[]{0.45\textwidth}
         \centering
         \includegraphics[width=\textwidth]{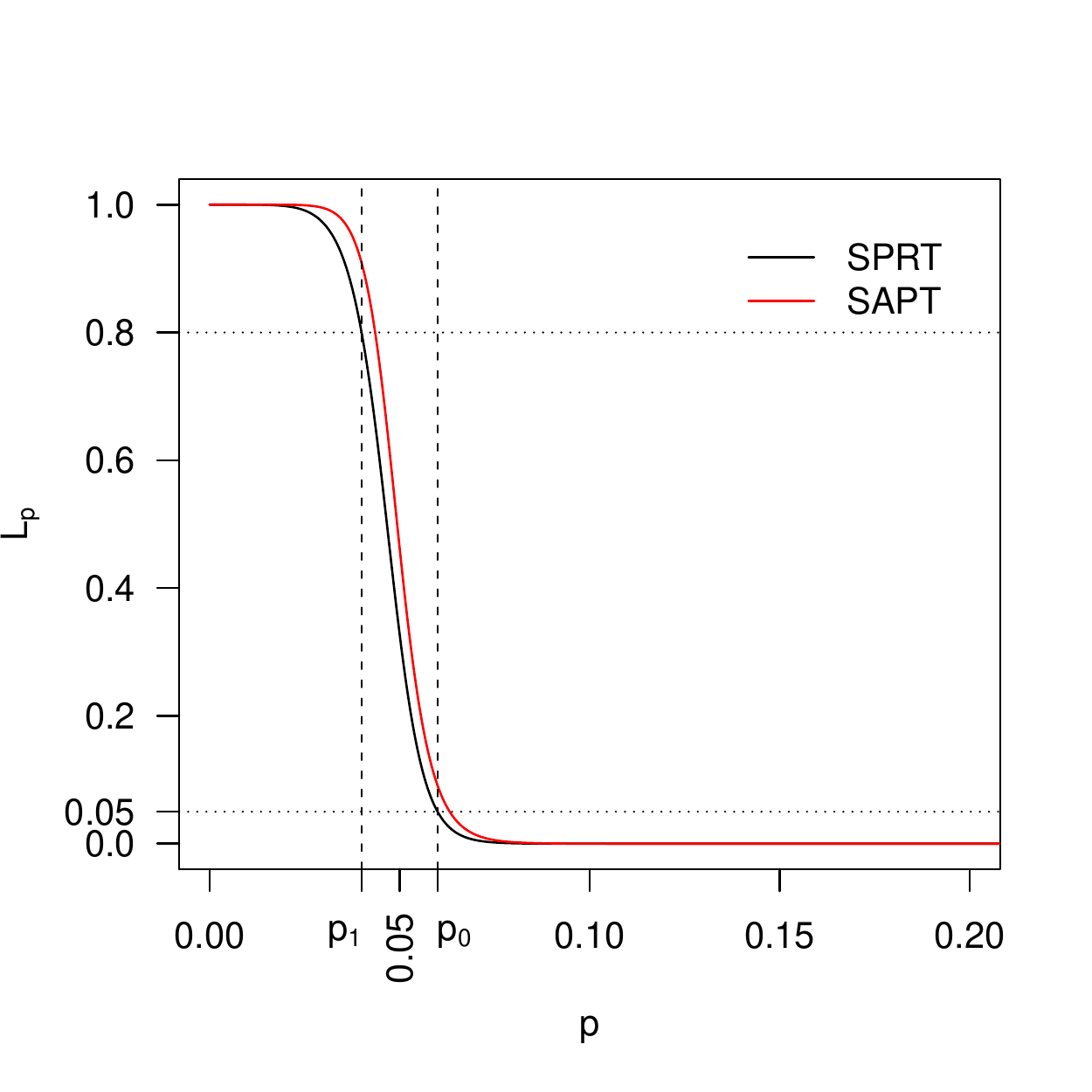}
         \caption{Power functions of SPRT and SAPT.}
         \label{fig:Power}
     \end{subfigure}
     \hfill
     \begin{subfigure}[]{0.45\textwidth}
         \centering
         \includegraphics[width=\textwidth]{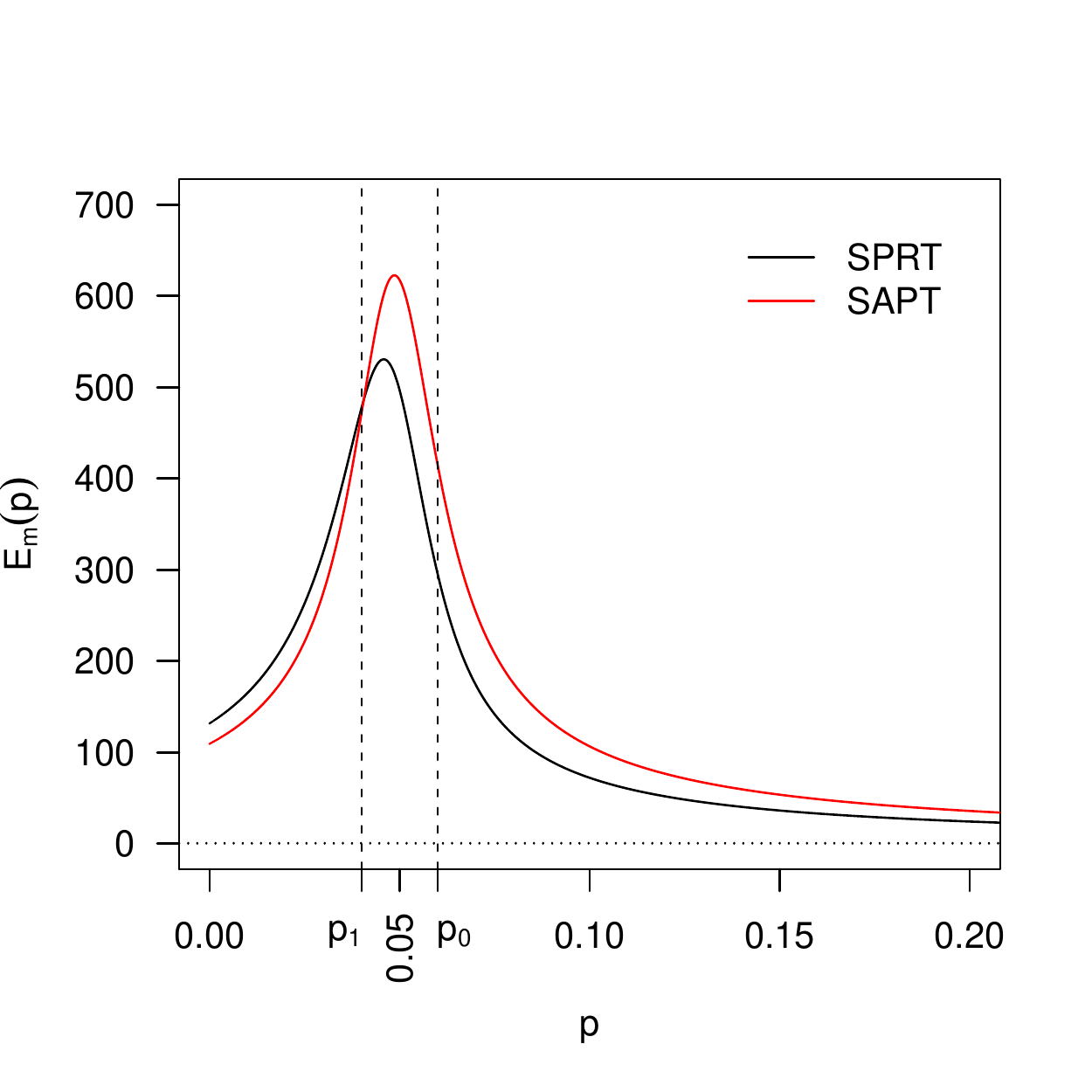}
         \caption{Expected number of permutations until termination of SPRT and SAPT.}
         \label{fig:Samplesize}
     \end{subfigure}
        \caption{Illustration of the power functions and the expected number of permutations until termination of SPRT with $(p_{0} = 0.06,\ p_{1} = 0.04,\ \alpha = 0.05,\ \beta = 0.20)^{\top}$ and SAPT with $(p_{0} = 0.06,\ p_{1} = 0.04,\ A = 0.1,\ B = 10)^{\top}$.}
        \label{fig:Power_Samp}
\end{figure}

The power function of the SPRT, which also applies to the SAPT, can be approximated by
\[
L_{p} = (1-A^{k})/(B^{k}-A^{k}),
\]
where $k$ satisfies the equation $p\, (p_{1}/p_{0})^{k} + (1-p)\, ((1-p_{1})/(1-p_{0}))^{k} = 1$  \citep[]{Wald1945, Lock1991}. Setting $p_{0} = 0.06$, $p_{1} = 0.04$, $\alpha = 0.05$, $\beta = 0.2$, $A = 0.1$ and $B = 10$, as outlined above, it can be seen from the respective power functions in Figure \ref{fig:Power} that the SAPT, unlike the SPRT, is not an exact size $\alpha$ test. While the rejection probability of $H_{0}{:}\; p = p_{0}$ at $p = p_{0}$ is $0.05$ for the SPRT, it is $0.09$ for the SAPT. However, exploiting the fact that $p$ follows a uniform distribution under $H_{0}$, \cite{Lock1991} derives the effective type-I error probability as the integral $\int_{0}^{1}L_{p}\ dp$, leading to values of $0.0465$ and $0.0498$ for the SPRT and the SAPT, respectively. In this respect, the SAPT makes better use of the permitted type-I error level $\alpha = 0.05$ and achieves a higher power when $p<p_{0}$ than the SPRT. 

The expected number of permutations until termination of the testing procedures is given by 
\[
E_{m}(p)=\frac{L_{p}\ log(B) + (1 - L_{p})\ log(A)}{p\ log(\frac{p_{1}}{p_{0}}) + (1 - p)\ log(\frac{1- p_{1}}{1- p_{0}})}.
\]
The SAPT is superior to the SPRT in terms of a lower expected number of permutations when $p \leq p_{1}$, that is when $H_{1}{:}\; p = p_{1}$ and more extreme alternatives are true (cf. Figure \ref{fig:Samplesize}). In line with this finding, the average expected number of permutations under $H_{0}$, again assuming a uniform distribution of $p$ and using the same values for $(p_{0}, p_{1}, A, B, \alpha, \beta)^{\top}$ as above, is $\int_{0}^{1}E_{m}(p)\ dp = 35.4$ for the SPRT and $43.6$ for the SAPT.   

\begin{figure}
\centering
\includegraphics[width=0.75\textwidth]{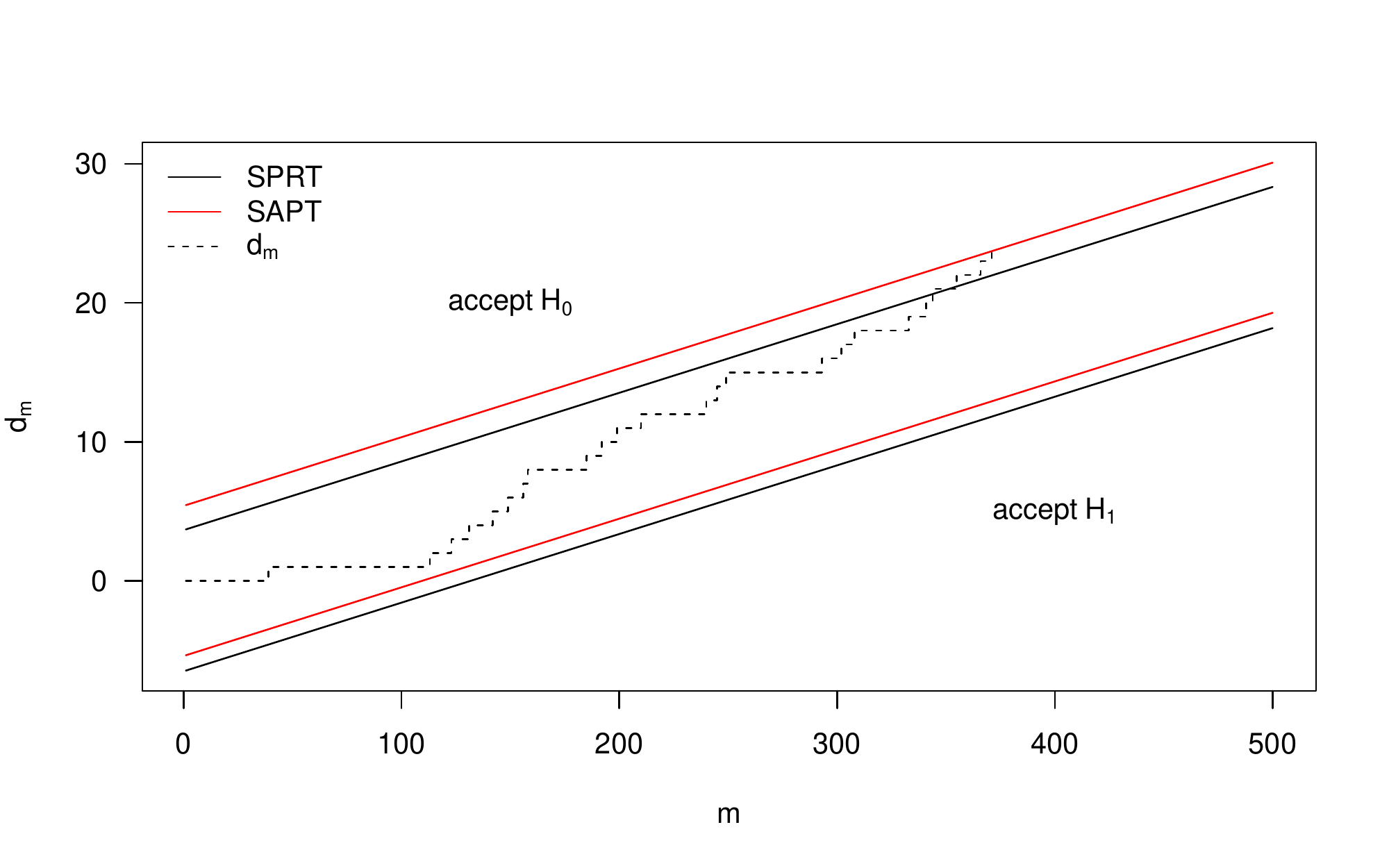}
\caption{Decision boundaries to accept $H_{0}$ or $H_{1}$ as a function of the number of permutations $m$ for SPRT with $(p_{0} = 0.06,\ p_{1} = 0.04,\ \alpha = 0.05,\ \beta = 0.20)^{\top}$ and SAPT with $(p_{0} = 0.06,\ p_{1} = 0.04,\ A = 0.1,\ B = 10)^{\top}$. A synthetic example of what a test result might look like has been added by sampling $I(u_{s} \geq u)$ in $d_{m} = \sum_{s=1}^{m}I(u_{s} \geq u)$ from a Bernoulli distribution with $p=0.05$ (dashed step function).}
\label{fig:Bounds}
\end{figure}

Another informative feature of the sequential tests is that the decision rules (\ref{eq:acceptsuccesses}) and (\ref{eq:rejectsuccesses}) for accepting one of the two hypotheses can be translated into linear decision boundaries for $d_{m}$, as shown in Figure \ref{fig:Bounds}. In line with the previous discussion about the power functions of the sequential tests and the expected number of permutations until termination of the testing procedures, this again shows that the SAPT tends to accept $H_{1}$ earlier and $H_{0}$ later than the SPRT. Also, the vertical distance between the decision boundaries of the SAPT is larger than that of the SPRT, which underlines the fact that the SPRT was introduced as the optimal or most efficient sequential test that requires the smallest expected number of permutations while controlling the type-I and type-II error probabilities \citep[]{Wald1945}.

\subsubsection{Sequential Monte Carlo p-values}

The SPRT and the SAPT can be used to accept either of the two hypotheses $H_{0}{:}\;p=p_{0}$ and $H_{1}{:}\; p=p_{1}$ prematurely, which has the advantage of considerably reducing the number of permutations compared to the modified and current most powerful test, which uses a fixed number of $M$ permutations. However, they also share the disadvantages that repeated applications can lead to inconsistent results, due to the random selection of permutations performed, and that they do not provide p-values \citep[]{Lock1991}. The latter issue can alternatively be solved by calculating a sequential Monte Carlo p-value \citep[]{Besag1991}. Therefore, the sequential p-value 
\begin{equation} \label{eq:pval}
\widehat{p} = 
  \begin{cases}
    d_{m}/m       & \quad \text{if } d_{m} = h \text{ and } m \leq M\\
    (d_{m} + 1)/(M+1)  & \quad \text{if } d_{m} < h \text{ and } m=M
  \end{cases}
\end{equation}
serves as an estimate of $p=1-F(u)$, where $F(\cdot)$ is the unknown null distribution of the random variable $U$ corresponding to the observed statistic $u$. The parameter $h$ is a fixed and predefined number of statistics $u_{s}$ that must exceed the statistic $u$ to satisfy the first condition in (\ref{eq:pval}) for early stopping. $\widehat{p}$ is a maximum likelihood estimate of $p$ in this case. A general recommendation for the choice of $h$ is $h=20$ or $h=10$ and even lower values have been suggested if computations are considered expensive \citep[]{Besag1991}. The sequential p-values are exact as $P(p \leq \alpha) = \alpha$ under $H_{0}$ when
\begin{equation} \label{eq:support}
\alpha \in \{1, h/(h+1), \ldots, h/(M-1), h/M, (h-1)/M, \ldots, 1/M\}.
\end{equation}
The set of values given in (\ref{eq:support}) is the support of the possible values of $\widehat{p}$ and has the advantageous property that the distribution of these values is denser for smaller values and thus in a range of values that is of major interest for the estimation of $p$. The precision of the estimate $\widehat{p}$, that is the standard error of $\widehat{p}$, is also directly related to $h$ as it is approximated by $p/\sqrt{h}$. With $h = 20$ or $h = 10$ it is therefore 22.4\% or 31.6\% of $p$. Finally, the average number of permutations needed in the computation of $\widehat{p}$ under $H_{0}$ is 
\[
E(m) \approx h + h\ log \{(M+\frac{1}{2})/(h+\frac{1}{2})\}.
\]
Again, with $h = 20$ or $h = 10$ and with $M = 500$, for example, $E(m)$ is 83.9 or 48.7. From the previous explanations it is evident that the choice of $h$ is a decision about the trade-off between computational feasibility and precision. 

Referring back to the SPRT and the SAPT, the expected number of permutations needed for early termination of these tests under $H_{0}$, and using the specifications as above, was determined to be 35.4 and 43.6, respectively. With $h=8$ a similar number of 40.6 permutations is expected for the computation of $\widehat{p}$ under $H_{0}$, with a respective precision of 35.4\% of $p$. It is worth noting at this point that any choice of $h \in \{8, 10, 20\}$ leads to an exact size-$\alpha$ test with $\alpha=0.05$, since it satisfies condition (\ref{eq:support}) when $M=500$. Moreover, the early stopping rule $d_{m} = h$ of the sequential p-value estimation can also be imagined as a horizontal line at height $h$ in Figure \ref{fig:Bounds}, which underlines its specific potential for early stopping when $d_{m}$ is large but still varies between the decision boundaries of the SPRT or the SAPT. The power of a test using $\widehat{p}$ is empirically investigated and compared with the SPRT and the SAPT in the simulation studies presented in this paper.

In summary, there are several options for stopping permutation tests at an early stage $m<M$. Using the decision rules (\ref{eq:earlystopp1}) and (\ref{eq:earlystopp2}), one can simply stop sampling permutations if the significance of the modified and current most powerful permutation test which uses $M$ permutations can already be established or cannot be reached any more at a stage $m<M$. Alternatively, the SPRT or SAPT can be applied using the decision rules (\ref{eq:acceptsuccesses}) and (\ref{eq:rejectsuccesses}) to minimise the expected number of permutations required until premature termination while controlling the type-I error probability and controlling or minimising the type-II error probability. Finally, early termination can be achieved by calculating sequential p-values using the case dependent equations (\ref{eq:pval}).

\subsection{Random Forest Permutation Importance Measure}

The most popular and widely used means of measuring the importance of a variable in a RF prediction model is the so-called 'Permutation Importance Measure' \citep[]{Strobl2009, Boulesteix2012, Hapfelmeier2013}. To simplify notation, it is generally referred to as 'VIMP' henceforth. Roughly said, the VIMP quantifies the relevance of a variable to prediction by comparing the prediction errors of a model when applied to the original data and to data where the variable of interest is randomly permuted \citep[cf.][for a more in-depth discussion of variable importance and different types of VIMP]{Hooker2021}. The intuitive idea behind the VIMP is that random permutation removes any potential and direct or indirect relationships of a variable to the outcome, which should lead to an increased prediction error of the model if the variable was relevant for the prediction. In the following, commonly used definitions are presented with slightly adapted notation to provide a formal description of a RF estimator and the VIMP \citep[cf.][]{Breiman2001, Ishwaran2019, Coleman2019}. 

A RF estimator of a function $g(X): \mathcal{X} \mapsto \mathcal{Y},\  x \mapsto g(x)$, which maps the sample space $\mathcal{X}$ of a covariate vector $X$ on the space $\mathcal{Y}$ of an outcome variable $Y$, can be formalized by 
\begin{equation} \label{eq:rf}
g(x,\Theta_{1},\ldots,\Theta_{T},L)=\frac{1}{T}\sum_{t=1}^{T}g(x,\Theta_{t},L).
\end{equation}
Here, $x$ is a realization of $X$, $L=\{(X_{1}, Y_{1}),\ldots,(X_{n},Y_{n})\} = (X,Y)$ is learning data of size $n$ with identically and independently distributed vectors $(X_{i},Y_{i})$, $i \in \{1,\ldots,n\}$, and $T$ is the total number of trees in the model. The randomization involved in the building process of the $t$-th tree model $g(\cdot,\Theta_{t},L)$, $t \in \{1,\ldots,T\}$, and in the computation of its VIMP, that is the sampling of data used to fit the tree, the sampling of variables used as split candidates, and the permutation of data used for the computation of the VIMP, is captured by the random quantity $\Theta_{t}$. With $T \to \infty$, a respective infinite RF estimator formulates to
\[
g(x,L)=\mathbb{E}_{\Theta}[g(x,\Theta,L)].
\]
Now, by using the difference as a contrast measure and by choosing an appropriate loss function $l(Y, g) \geq 0$, for example a squared error loss in case of a continuous outcome or the Brier-score in case of a categorical outcome, the VIMP of a variable $X_{j}$ in the $t$-th tree predictor $g$ defines to
\begin{equation} \label{eq:vimptree}
VI(X_{j},\Theta_{t},L)=\frac{\sum_{i \in L^{*}(\Theta_{t})} l(Y_{i}, g(\tilde{X}_{i},\Theta_{t},L))}{|L^{*}(\Theta_{t})|} - \frac{\sum_{i \in L^{*}(\Theta_{t})} l(Y_{i}, g(X_{i},\Theta_{t},L))}{|L^{*}(\Theta_{t})|},
\end{equation}
where $X_{i}$ is the $i$-th observation or row vector in $X$. $L^{*}(\Theta_{t})$ is the $t$-th so-called 'out-of-bag' (OOB) sample of $L$, that is the fraction of data which is not used to build the $t$-th tree as coded by $\Theta_{t}$. $L^{*}(\Theta_{t})$ is of size $|L^{*}(\Theta_{t})|$. The tilde indicates that $X_{j}$ is permuted in the addressed OOB data.

Finally, the VIMP of $X_{j}$ in the whole RF model is given by
\begin{equation} \label{eq:vimp}
VIMP(X_{j}) = VI(X_{j},\Theta_{1},\ldots,\Theta_{T},L)=\frac{1}{T}\sum_{t=1}^{T}VI(X_{j},\Theta_{t},L).
\end{equation}
And again, a respective infinite VIMP is defined as
\begin{equation} \label{eq:vimpinf}
VIMP_{\Theta}(X_{j}) = VI(X_{j},L) = \mathbb{E}_{\Theta}[VI(X_{j},\Theta,L)]
\end{equation}
with $T \to \infty$ and conditional on the learning data $L$.

\subsection{Permutation testing of VIMP}

Sampling methods like the bootstrapping, jackknife estimation and empirical investigations have been used to approximate the distribution of VIMP, to estimate its variance and to establish respective hypothesis tests \citep[]{Ishwaran2019, Janitza2018}. Other sophisticated analytical approaches to formal hypothesis testing of variable importance under regularity conditions have been found to be computationally expensive or even practicably unfeasible \citep[][]{Mentch2016, Mentch2017}.

For a similar rationale, a permutation test of the null hypothesis 
\[
H_{0}{:}\; VIMP_{L,\Theta}(X_{j}) = VI(X_{j}) = \mathbb{E}_{L}\mathbb{E}_{\Theta}[VI(X_{j},\Theta,L)] = 0
\]
was proposed in our own earlier work to circumvent the necessity of distributional assumptions about VIMP in a pragmatic approach \citep[]{Hapfelmeier2013, Hapfelmeier2014}. Here the estimate $\widehat{VIMP}(X_{j})$, which is estimated by replacing $\Theta$, $L$ and $g$ by the realizations $\theta$, observed data $(x,y)$, and the fitted predictors $\widehat{g}$, takes the place of the statistic $u$ used for permutation testing. The respective statistics $u_{s}$ are given by multiple estimates $\widehat{VIMP}(\tilde{X}_{j})$ which are computed from repeated fits of the RF model to data $(\tilde{x},y)$, where in each permutation step, $x_{j}$ is replaced by the randomly permuted $\tilde{x}_{j}$. The test procedure then follows as described in section \ref{sec:permtest} and as outlined by the pseudocode of the algorithm given in Table \ref{table:vimptest}. Refitting a RF model and recomputing the VIMP in potentially many permutation steps  is of course computationally expensive. However, the approach generalizes well to any type of predictive model if using a suitable statistic $u$. It is therefore the first approach proposed in this work to be used in combination with the sequential permutation tests and p-value estimation described in section \ref{sec:permtest}.

\begin{table}
\centering
\begin{tabular}{l} 
 \hline
 1. Fit a RF model $\widehat{g}(\cdot,\theta_{1},\ldots,\theta_{T},(x,y))$. \\
 2. Estimate $u = \widehat{VIMP}(X_{j}) = VI(x_{j},\theta_{1},\ldots,\theta_{T},(x,y))$. \\ 
 3. for $s \in \{1,\ldots,M$\} \\
 - Randomly permute $x_{j}$ to make $\tilde{x}_{j}$. \\
 - Replace $x_{j}$ in $(x,y)$ by $\tilde{x}_{j}$ to make $(\tilde{x},y)$. \\
 - Fit a RF model $\widehat{g}(\cdot,\tilde{\theta}_{1},\ldots,\tilde{\theta}_{T},(\tilde{x},y))$. \\
 - Estimate $u_{s} = \widehat{VIMP}(\tilde{X}_{j}) = VI(\tilde{x}_{j},\tilde{\theta}_{1},\ldots,\tilde{\theta}_{T},(\tilde{x},y))$. \\
 4. Return $p= \frac{1}{M} \sum_{s = 1}^{M} I(u_{s} \geq u)$. \\ 
 \hline
\end{tabular}
\caption{Pseudocode for a general permutation test of the VIMP of a variable $X_{j}$. The tilde in $\tilde{\theta}_{t}$ indicates that the respective RF models are refit to the data $(\tilde{x},y)$.}
\label{table:vimptest}
\end{table}

A second approach considered in this work has been suggested specifically for RF, though it can be applied to any kind of ensemble learner \citep[]{Coleman2019}. It makes use of the two-sample permutation test for equality of distribution, relying on the exchangeability of trees in a RF model. In this sense, the predictive relevance of a variable $X_{j}$ is tested by fitting two RF models $\widehat{g}(\cdot,\theta_{1},\ldots,\theta_{T},(x,y))$ and $\widehat{g}(\cdot,\tilde{\theta}_{1},\ldots,\tilde{\theta}_{T},(\tilde{x},y))$ to the original data and to data where $x_{j}$ has been replaced once by the randomly permuted $\tilde{x}_{j}$, respectively. The tilde in $\tilde{\theta}_{t}$ therefore indicates that the respective RF model is fit to the data $(\tilde{x},y)$. The two samples compared are then represented by vectors containing the respective predictions $\widehat{g}(x^{*},\theta_{t},(x,y))$ and $\widehat{g}(x^{*},\tilde{\theta}_{t},(\tilde{x},y))$ of some independent test data $x^{*}$, including outcomes $y^{*}$, made by the $t=1,\ldots,T$ individual trees of each of the two RF models. For permutation testing it has been suggested to use the difference of the mean squared errors of the two RF models as the statistic $u$ \citep[]{Coleman2019}. Following the principle of a two-sample permutation test, the statistics $u_{s}$ are then calculated based on a random permutation of the trees, that is, after the trees have been randomly assigned in equal numbers to one of the RF models. This is essentially a random permutation of the pooled trees leading to an altered composition of the RF models. The test procedure is then again carried out as described in section \ref{sec:permtest}. The underlying rationale is that the mean squared errors of the RF models are not affected by the permutation of trees if a variable $X_{j}$ is not relevant to the prediction. The advantage of this approach is that only two RF models need to be fit, which considerably reduces the computational effort.

A heuristic variant inspired by the second approach is proposed in this paper. It uses the estimated VIMPs $\widehat{VI}(x_{j},\theta_{t},(x,y))$ and $\widehat{VI}(\tilde{x}_{j},\tilde{\theta}_{t},(\tilde{x},y))$ of the variables $X_{j}$ and $\tilde{X}_{j}$ in the $t=1,\ldots,T$ trees of each of the two RF models $\widehat{g}(\cdot,\theta_{1},\ldots,\theta_{T},(x,y))$ and $\widehat{g}(\cdot,\tilde{\theta}_{1},\ldots,\tilde{\theta}_{T},(\tilde{x},y))$ as the compared samples. Thereby it suffices, and is therefore computationally more efficient, to use the statistic $u = \widehat{VIMP}(X_{j}) = \widehat{VI}(x_{j},\theta_{1},\ldots,\theta_{T},(x,y))$ and the statistics 
\[
u_{s} = \frac{1}{T}\sum_{t=1}^{T} r_{t} \ \widehat{VI}(x_{j},\theta_{t},(x,y)) + \frac{1}{T}\sum_{t=1}^{T} \tilde{r}_{t} \  \widehat{VI}(\tilde{x}_{j},\tilde{\theta}_{t},(\tilde{x},y)) 
\]
in the respective two-sample permutation test. Herein, the two vectors $r \in \{0,1\}$ and $\tilde{r} \in \{0,1\}$ are of length $T$ with $\sum_{t=1}^{T} r_{t} + \tilde{r}_{t} = T$. They indicate which trees have been randomly sampled from both RF models to compute each $u_{s}$. Respective pseudocode of the algorithm is given in Table \ref{table:vimptest2}. The approach can again be used in combination with the sequential permutation tests and p-value estimation described in section \ref{sec:permtest}.

\begin{table}
\centering
\begin{tabular}{l} 
 \hline
 1. Fit a RF model $\widehat{g}(\cdot,\theta_{1},\ldots,\theta_{T},(x,y))$. \\
 2. Randomly permute $x_{j}$ to make $\tilde{x}_{j}$. \\
 3. Replace $x_{j}$ in $(x,y)$ by $\tilde{x}_{j}$ to make $(\tilde{x},y)$. \\
 4. Fit a RF model $\widehat{g}(\cdot,\tilde{\theta}_{1},\ldots,\tilde{\theta}_{T},(\tilde{x},y))$. \\
 5. for $t \in \{1, \ldots, T\}$ \\
 - Estimate $\widehat{VI}(x_{j},\theta_{t},(x,y))$. \\
 - Estimate $\widehat{VI}(\tilde{x}_{j},\tilde{\theta}_{t},(\tilde{x},y))$. \\
 6. Calculate $u = \widehat{VIMP}(X_{j}) = \widehat{VI}(x_{j},\theta_{1},\ldots,\theta_{T},(x,y))=\frac{1}{T}\sum_{t=1}^{T} \widehat{VI}(x_{j},\theta_{t},(x,y))$. \\ 
 7. for $s \in \{1,\ldots,M$\} \\
 - Randomly sample a vector $v$ of length $T$ from $\{\widehat{VI}(x_{j},\theta_{t},(x,y))\} \cup \{\widehat{VI}(\tilde{x}_{j},\tilde{\theta}_{t},(\tilde{x},y))\}$ \\ without replacement. \\
 - Calculate $u_{s} = \frac{1}{T}\sum_{t=1}^{T} v_{t}$. \\
 4. Return $p= \frac{1}{M} \sum_{s = 1}^{M} I(u_{s} \geq u)$. \\
 \hline
\end{tabular}
\caption{Pseudocode for a two-sample permutation test of the VIMP of a variable $X_{j}$. The tilde in $\tilde{\theta}_{t}$ indicates that the respective RF model is fit to the data $(\tilde{x},y)$.}
\label{table:vimptest2}
\end{table}

Although this case is not addressed in the present work, it should be mentioned that all of the aforementioned approaches can generally be extended to test the mutual importance of multiple variables by permuting the values of a respective set of variables simultaneously in the data.

In summary, a total of eight sequential and two non-sequential testing approaches can be considered. The general permutation test (cf. Table \ref{table:vimptest}) or the two-sample permutation test (cf. Table \ref{table:vimptest2}) can be combined with the SPRT or the SAPT (using the decision rules (\ref{eq:acceptsuccesses}) and (\ref{eq:rejectsuccesses})), with the computation of sequential p-values (using the case-distinctive definition (\ref{eq:pval}); henceforth called PVAL), with early stopping when a decision is already certain (using the decision rules (\ref{eq:earlystopp1}) and (\ref{eq:earlystopp2}); henceforth called CERTAIN) or with the modified and current most powerful test which does not use early stopping (henceforth called COMPLETE). The following naming of the methods is also shown in table \ref{table:methodnames} for the sake of clarity.

\begin{table}
\centering
\begin{tabular}{lll} 
 \hline
& permutation test & \\
sequential method & general & two-sample \\ \hline
SPRT & SPRT & SPRT\_TS \\
SAPT & SAPT & SAPT\_TS \\
p-value & PVAL & PVAL\_TS \\
certain stopping & CERTAIN & CERTAIN\_TS \\
none & COMPLETE & COMPLETE\_TS \\
 \hline
\end{tabular}
\caption{Naming of the sequential testing approaches as defined by the combination of the type of permutation test (columns) and method used for sequential testing (rows).}
\label{table:methodnames}
\end{table}

\section{Simulation Studies}

The properties of the methods are examined in terms of type-I error probability, type-II error probability (or power) and the number of permutations required until premature termination using two types of simulation studies. For all analyses, unless explicitly stated otherwise, the maximum number of permutations was set to $M=500$ for the permutation tests and $h=8$ was set for the PVAL method. A direct comparison of the methods was made possible by applying COMPLETE to produce a sequence of values $\{d_{m}\}$, $m \in \{1,\ldots,M\}$. The decision rules of the other methods were then applied to this sequence to determine if and when they would have terminated prematurely. Consequently, the methods are applied to the same simulated data with the same permutations of variables. The same applies to the following application studies in section \ref{sec:appstud}.

The setting of the first simulation study has been used before to examine the type-I error probability and the power of COMPLETE \citep[]{Hapfelmeier2013}. It presents a classification problem where the binary outcome $Y$ follows a Bernoulli distribution $B(1, \pi)$, that is
\[
Y \sim B(1, \pi) \ \text{ with } \ \pi = P(Y=1|X=x) = e^{x^{\top} \beta } / (1 + e^{x^{\top} \beta }).
\]

For the linear predictor $x^{\top} \beta$, the parameters $\beta = (k,k,k,1,0,0)^{\top}$ were chosen to cover six cases of increasingly informative and non-informative variables $X$, corresponding to $k \in \{0, 0.125, 0.250, \ldots, 1\}$. The variables $X$ follow a multivariate standard normal distribution with $Cov(X_{2}, X_{4}) = Cov(X_{3}, X_{5}) = 0.5$ and no correlations between other variables to generate two pairwise correlations. In this regard, the so called 'conditional permutation importance' has been suggested to quantify the partial relevance of a predictor variable in the RF prediction model that is not affected by correlation of this variable to other relevant predictor variables \citep[]{Strobl2008, Debeer2020}. However, the present paper is concerned with the original definition of VIMP, which measures the marginal relevance of a variable to prediction. Thereby, a variable can also be relevant for the prediction because it correlates with an informative variable. In order to avoid redundancy, the present study focuses on the variables $X_1$, $X_2$, and $X_5$. Here $X_1$ represents an increasingly informative variable that is uncorrelated with the other variables, $X_2$ represents an increasingly informative variable that is correlated with another informative variable, and $X_5$ represents a non-informative variable that gains relevance through correlation with an increasingly informative variable.

For this analysis, the usual default value of $ntree = 500$ trees in the RF, fit with the $ctree$ function of the R package $party$, was chosen. Likewise, the default value $nperm = 1$ was used to calculate the VIMP by the function $permimp$ of the $permimp$ R package \citep[]{Debeer2020}, which means that no averaged VIMP is calculated over multiple permutations of the OOB values of a variable. The sample size is set to $n=100$ as in the original simulation study \citep[]{Hapfelmeier2013}. The number of randomly sampled variables to determine an optimal split was chosen to be the square root of the total number of predictor variables, which is a common recommendation for classification problems \citep[]{Liaw2002}. In the present case this results to $mtry = 3$. 

The second simulation study is meant to cover more complex non-linear and interacting associations of some informative predictor variables with the outcome in a regression problem. The well-known Friedman 1 simulation has been widely used to assess the performance of machine learning methods like RF and associated estimators \citep[]{Friedman1991, Breiman1996, Ishwaran2019, Leisch2021}. In Friedman 1 the outcome is given by 
\[
y=10\, sin(\pi x_{1} x_{2}) + 20(x_{3}-0.5)^{2} + 10x_{4} + 5x_{5} +\epsilon
\]
with $\epsilon \sim N(0, 1)$. In total, there are ten variables $X_{1}, \ldots, X_{10}$, five informative variables and five non-informative variables, each following a uniform distribution $U(0, 1)$. While all variables play an important role in this simulation, only $X_{1}$, $X_{3}$ and $X_{6}$ are considered further in the following. The investigation of the other variables would either be redundant, for example regarding the additional non-informative variables and $X_{2}$, or less interesting, for example regarding the linear effects of $X_{4}$ and $X_{5}$. 

Again, $ntree = 500$, $nperm = 1$, and $n=100$ were chosen for this analysis. Furthermore, $mtry = 4$ was chosen as this corresponds to one third of the total number of predictor variables, which is a common recommendation for regression problems \citep[]{Liaw2002}. The data was sampled using the function $mlbench.friedman1$ of the $mlbench$ R package \citep[]{Leisch2021}. 

For both simulation settings, 1000 datasets were generated.

\section{Results}
\subsection{Simulation study I}

\begin{figure}
\centering
\includegraphics[width=\textwidth]{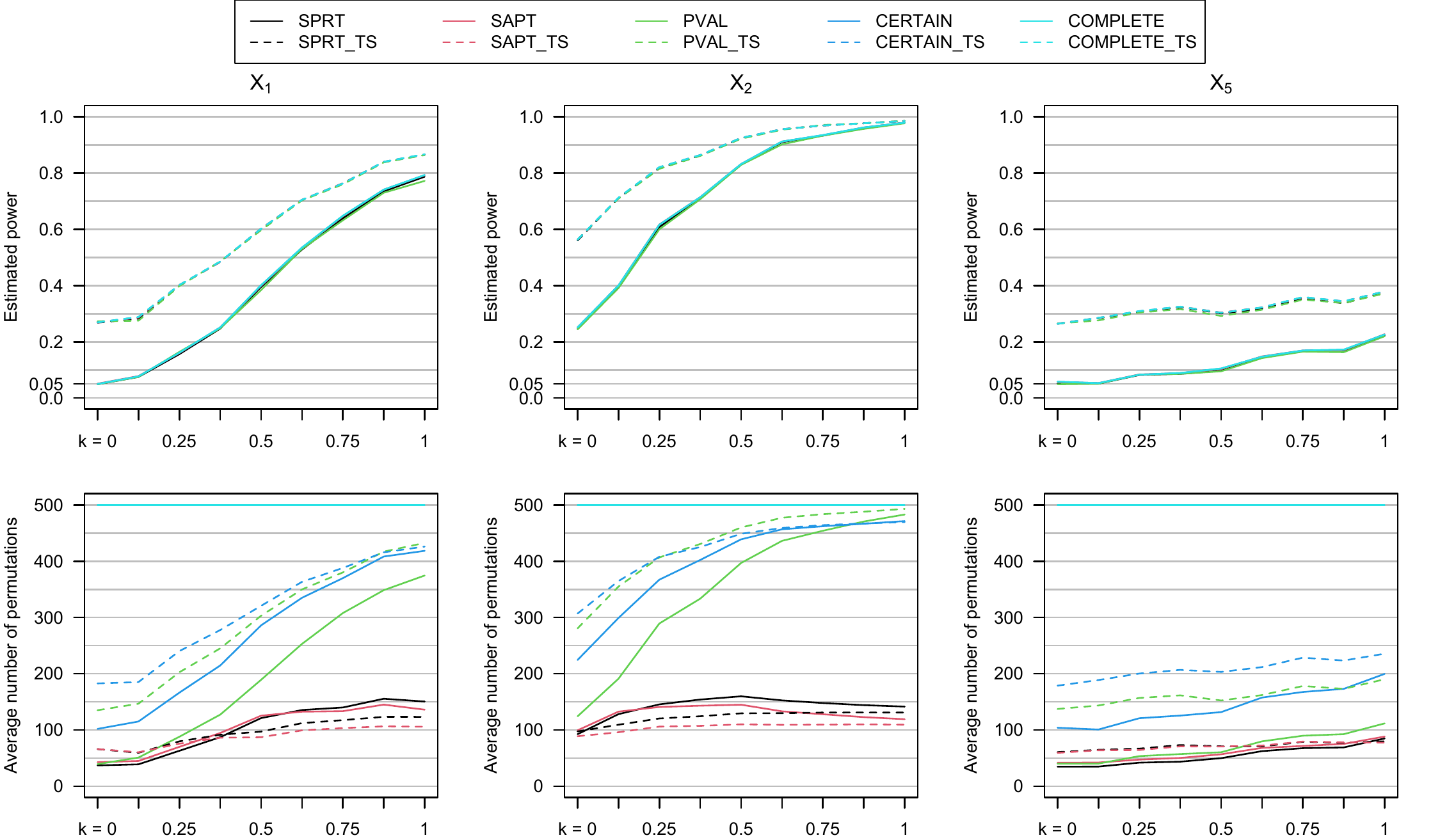}
\caption{Plots of estimated power (= relative rejection frequency of $H_{0}$) and average number of permutations by increasing effect size $s$ for the variables $X_{1}$, $X_{2}$ and $X_{5}$ in the first simulation study.}
\label{fig:StudyI}
\end{figure}

Results of the first simulation study are shown in Figure \ref{fig:StudyI}. The type-I error probabilities of the test methods are estimated by the relative rejection frequencies of $H_{0}$ when $X_{1}$ and $X_{5}$ are not informative, that is when $k=0$. For SPRT, SAPT and PVAL, using general permutation tests, these estimates range between 0.049 and 0.051 for $X_{1}$ and between 0.049 and 0.057 for $X_{5}$. Thus, no concerning deviations from a type I error control at a nominal significance level of $\alpha = 0.05$ were observed. Using COMPLETE and CERTAIN as a benchmark, the respective values were 0.050 and 0.058 for $X_{1}$ and $X_{5}$. In Section \ref{sec:permtest}, the expected number of permutations until early termination under $H_{0}$ was calculated to be 35.4, 43.6 and 40.6 for SPRT, SAPT and PVAL. In the present simulation study similar average values of 37.0, 42.7 and 39.2 were observed for $X_{1}$, respectively. Corresponding and similar values for $X_{5}$ were 34.7, 41.7 and 40.1. 

The respective estimates of the type-I error probabilities ranged between 0.264 and 0.273 for the two-sample permutation tests. This equals a five- to six-fold increase above the nominal significance level of $\alpha = 0.05$, disqualifying the methods from further considerations of their power when informative variables are taken into account.

The general permutation tests achieve a similar estimated power with respect to $X_{2}$ as well as $X_{1}$ and $X_{5}$ at $k>0$. The power of SPRT, SAPT and PVAL averaged 98.2\%, 99.9\% and 97.8\% of the power of COMPLETE and CERTAIN, respectively, for all simulated cases. 

With respect to the average number of permutations, significant savings can already be observed in CERTAIN. When $X_{1}$ and $X_{5}$ are not informative, that is when $k=0$, PVAL uses as few permutations as SPRT and SAPT. However, the average number of permutations increases with increasing $k>0$ for CERTAIN and PVAL, which is consistent with the definition of their decision rules (cf. Section \ref{sec:permtest}). The lowest average number of permutations is observed for SPRT and SAPT. These results also confirm that with increasing evidence for the respective hypothesis, SPRT can accept a true $H_{0}$ earlier, while SAPT can accept a true $H_{1}$ earlier, even though the differences seem to be small in the present setting. Both methods require lower numbers of permutations when the variables are non-informative or strongly informative, and a higher number of permutations in between. These findings are consistent with the power functions, expected number of permutations under $H_{0}$ and decision bounds shown in Figure \ref{fig:Power_Samp} and Figure \ref{fig:Bounds}.

\subsection{Simulation study II}

\begin{figure}
\centering
\includegraphics[width=\textwidth]{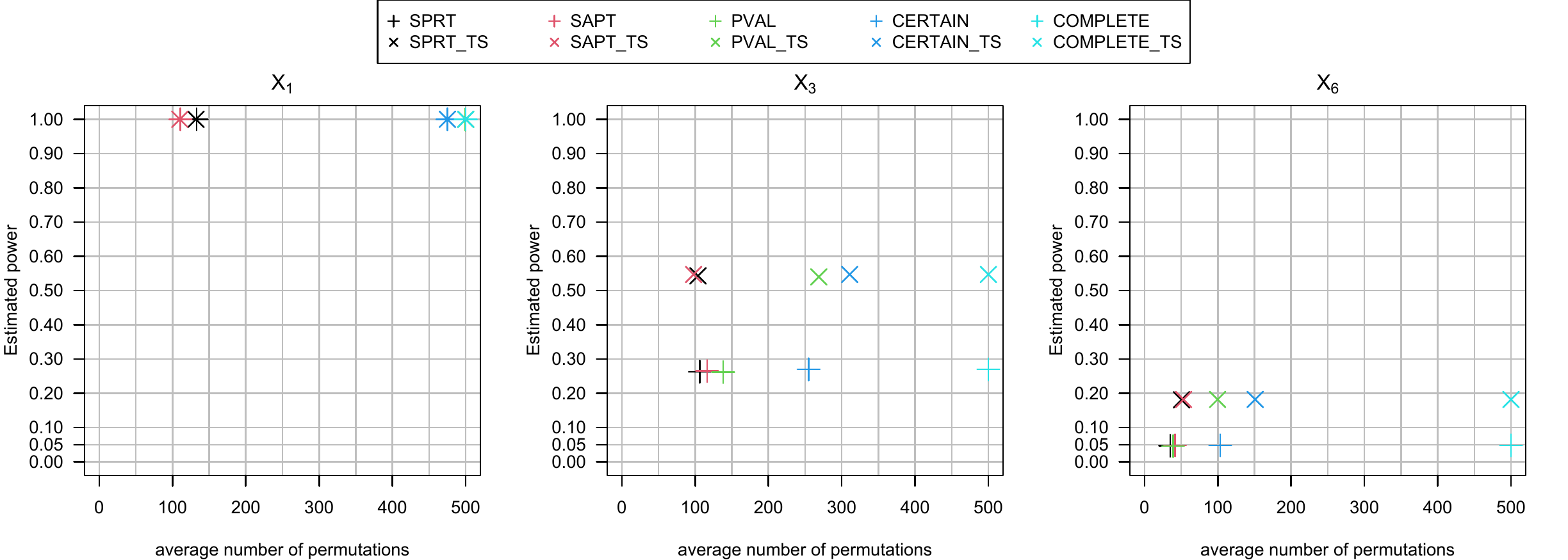}
\caption{Estimated power (= relative rejection frequency of $H_{0}$) and average number of permutations performed for the variables $X_{1}$, $X_{3}$ and $X_{6}$ in the second simulation study of the Friedman 1 regression problem.}
\label{fig:Friedman1}
\end{figure}

Results of the second simulation study based on the Friedman 1 regression problem are displayed in Figure \ref{fig:Friedman1}. Concerning the non-informative variable $X_{6}$, the estimated type-I error probability was well controlled at the nominal level $\alpha = 0.05$, taking values between 0.046 and 0.048 for the general permutation tests. The average number of permutations performed for SPRT, SAPT and PVAL were 35.0, 41.6 and 38.9 and were close to the expected numbers of permutations under $H_{0}$ of 35.4, 43.6 and 40.6 as outlined in section \ref{sec:permtest}. However, for the two-sample permutation test, the estimated type-I error probability exceeded the nominal level $\alpha = 0.05$ by a factor of about 3.6, ranging between 0.181 and 0.182. The higher power of the two-sample permutation tests in terms of accepting $H_{1}$ for the informative variable $X_{3}$ is therefore not considered useful and will not be discussed further in the following. 

The estimated power of COMPLETE and CERTAIN is 0.270 with respect to variable $X_{3}$. Similar estimates (relative to COMPLETE and CERTAIN) of 0.263 (97.4\%), 0.265 (98.1\%) and 0.262 (97.0\%) were obtained for SPRT, SAPT and PVAL, respectively, at considerably reduced average numbers of 106.4, 116.2 and 138.2 permutations. CERTAIN also required only 254.8 permutations on average. Accepting $H_{1}$ for the informative variable $X_{1}$ was easy for any of the methods leading to estimates of the power $\geq 0.999$. Considerably reduced average numbers of 132.9 and 110.8 permutations are observed only for SPRT and SAPT, again showing that SAPT is able to accept $H_{1}$ earlier when there is strong evidence in favour of $H_{1}$. For PVAL and CERTAIN the average number remained high at 499.5 and 475.1. The reason for this is that early stopping of PVAL relies on observing a certain number of statistics $u_{s} \geq u$, or equivalently higher values of $d_{m}$. With a strong predictor like $X_{1}$ such observations are evidently not frequent enough to lead to early stopping. Also, CERTAIN cannot accept $H_{1}$ before 95\% of the permutations have been performed (cf. decision rule (\ref{eq:earlystopp2}) in Section \ref{sec:permtest}). 

\section{Application Studies}
\label{sec:appstud}

The variability of results, which is due to the non-deterministic nature of the testing procedures \citep[][]{Lock1991}, and the randomness involved in model fitting and computation of the VIMP, is investigated in an application to two real data sets.

\subsection{Pima Indians Diabetes Database}

The first real data classification problem is based on the Pima Indians Diabetes Database, a classic provided by the UCI Machine Learning Repository \citep[][]{Dua2019}. This application study is intended to create comparability to many other studies that also use these data. The binary outcome is the diabetes status (yes/no) of $n=768$ Pima Indian women. The eight predictor variables are the number of pregnancies, plasma glucose concentration measured by a glucose tolerance test, diastolic blood pressure (mmHg), triceps skin fold thickness (mm), 2-Hour serum insulin (mu U/ml), BMI, diabetes pedigree function and age (years). Missing values are present in plasma glucose concentration (0.7\%), diastolic blood pressure (4.6\%), triceps skin fold thickness (29.6\%), 2-Hour serum insulin (48.7\%) and BMI (1.4\%) \citep[]{Pearson2006, Hapfelmeier2014b}. In the present analysis, two measures are taken to examine the variability of the methods' decisions to accept one of the two hypotheses and of the estimated p-values. One measure is to conduct a complete case analysis, which reduces the sample size to 392 (51.0\%). The other measure is to focus on three variables with the lowest estimated VIMP, that is diastolic blood pressure, triceps skin fold thickness and diabetes pedigree function \citep[cf.][]{Hapfelmeier2014b}. With more observations available for analysis and with stronger predictor variables, the decisions of the methods showed almost no variability, which is why these cases are not of interest for further consideration.

To obtain a benchmark for the results of hypothesis testing, the COMPLETE method is applied 20 times, using $M=10000$ permutations for the permutation tests, $nperm=3$ permutations for the calculation of an averaged estimate of VIMP, and $ntree=1000$ trees for the RF model. The respective performance of all methods is examined as they are applied 100 times to the data with $M=500$ and two different settings each for $ntree \in \{500, 1000\}$ and $nperm \in \{1, 3\}$. Following the aforementioned recommendation for classification problems, the number of variables used to determine an optimal split is set to $mtry=3$. The data itself is provided by the $mlbench$ R package \citep[][]{Leisch2021}. Therein, the data is called PimaIndianDiabetes2.

\begin{figure}
     \centering
     \begin{subfigure}[]{0.35\textwidth}
         \centering
         \includegraphics[width=\textwidth]{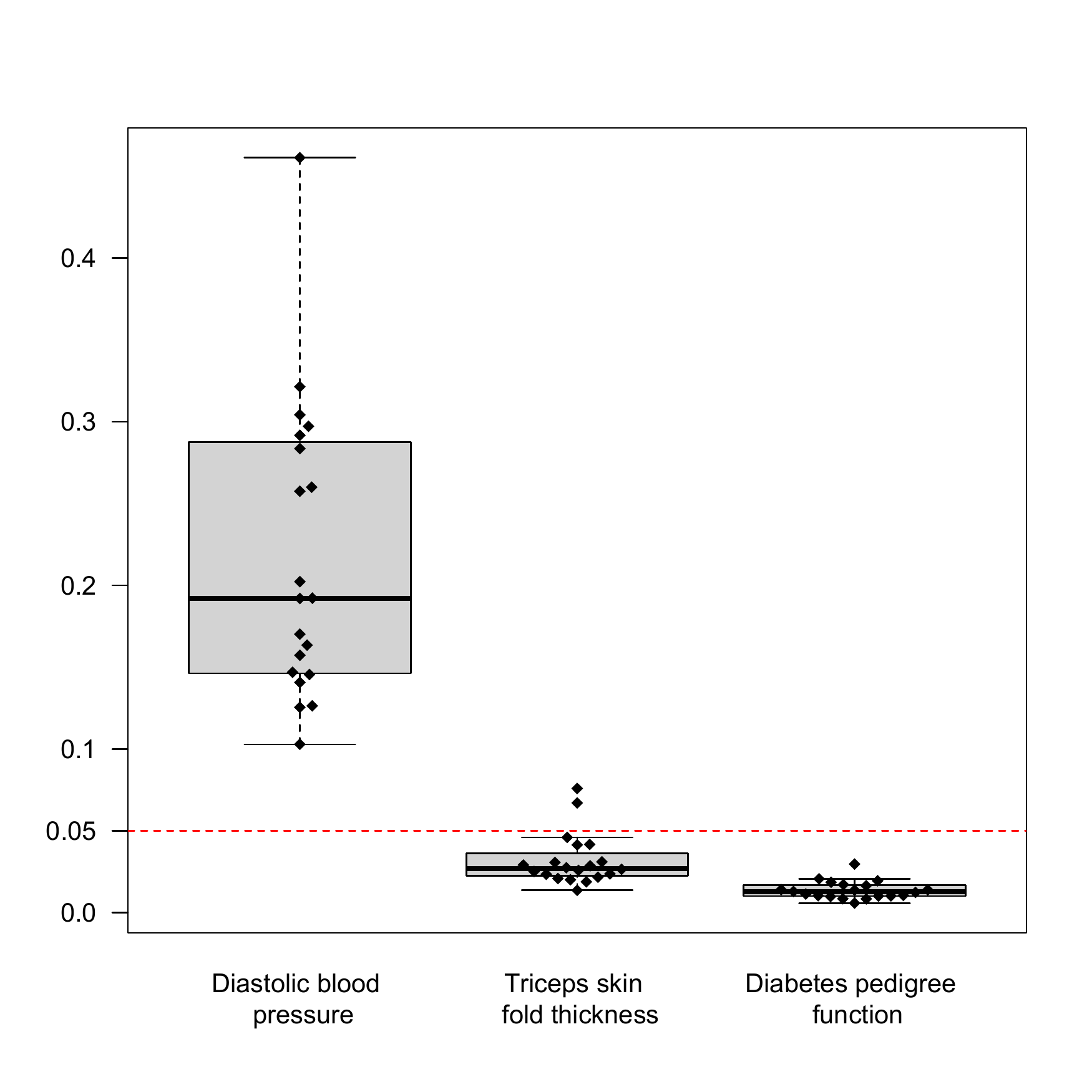}
         \caption{Distribution of p-values estimated by COMPLETE ($M=10000$).}
         \label{fig:Pima1}
     \end{subfigure}
     \hspace{0.075\textwidth} 
     \begin{subfigure}[]{0.35\textwidth}
         \centering
         \includegraphics[width=\textwidth]{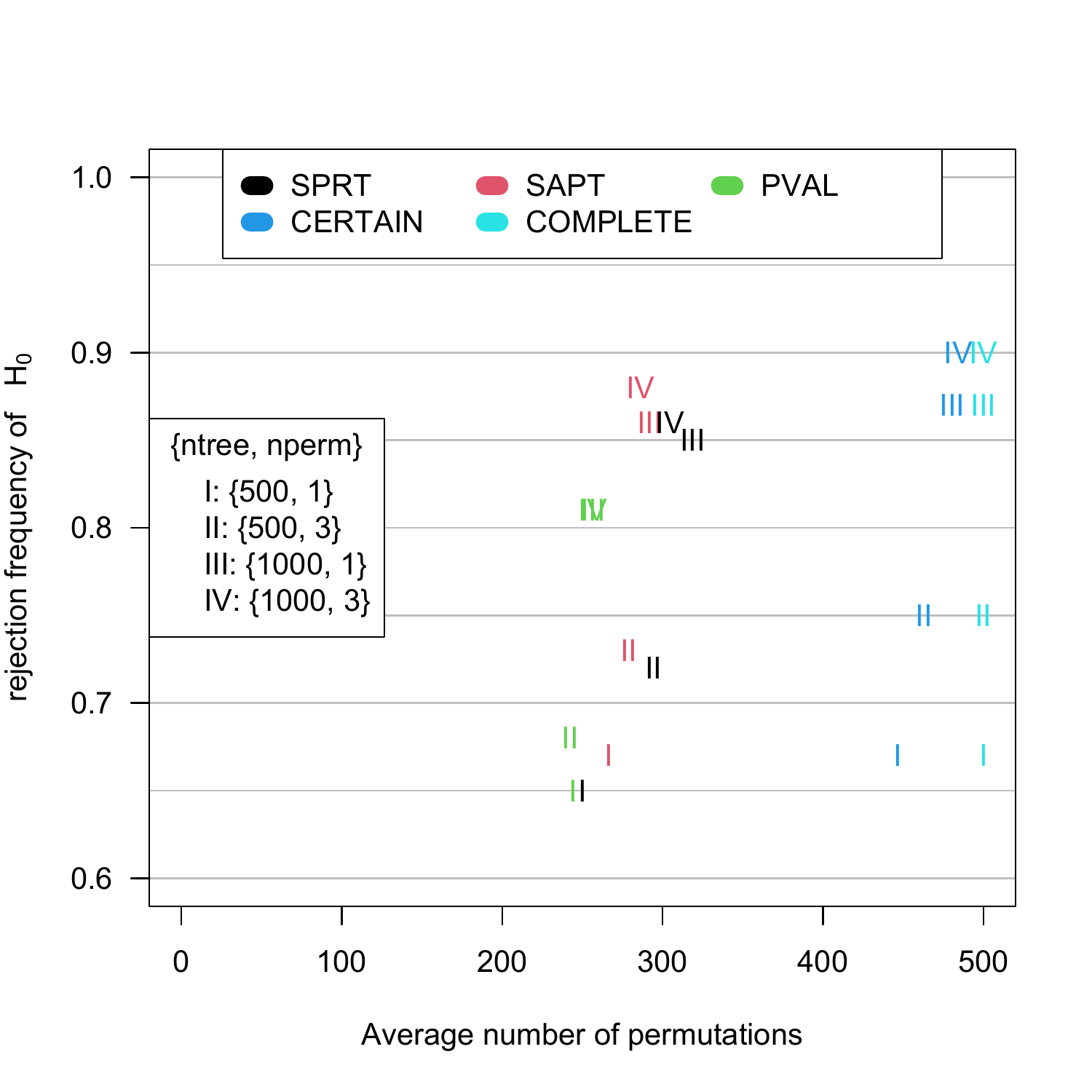}
         \caption{Relative rejection frequency of $H_{0}$ for triceps skin fold thickness by average number of permutations ($ntree \in \{500, 1000\}$, $nperm \in \{1, 3\}$).}
         \label{fig:Pima2}
     \end{subfigure}
     \hfill \\
     \begin{subfigure}[]{0.80\textwidth}
         \centering
         \includegraphics[width=\textwidth]{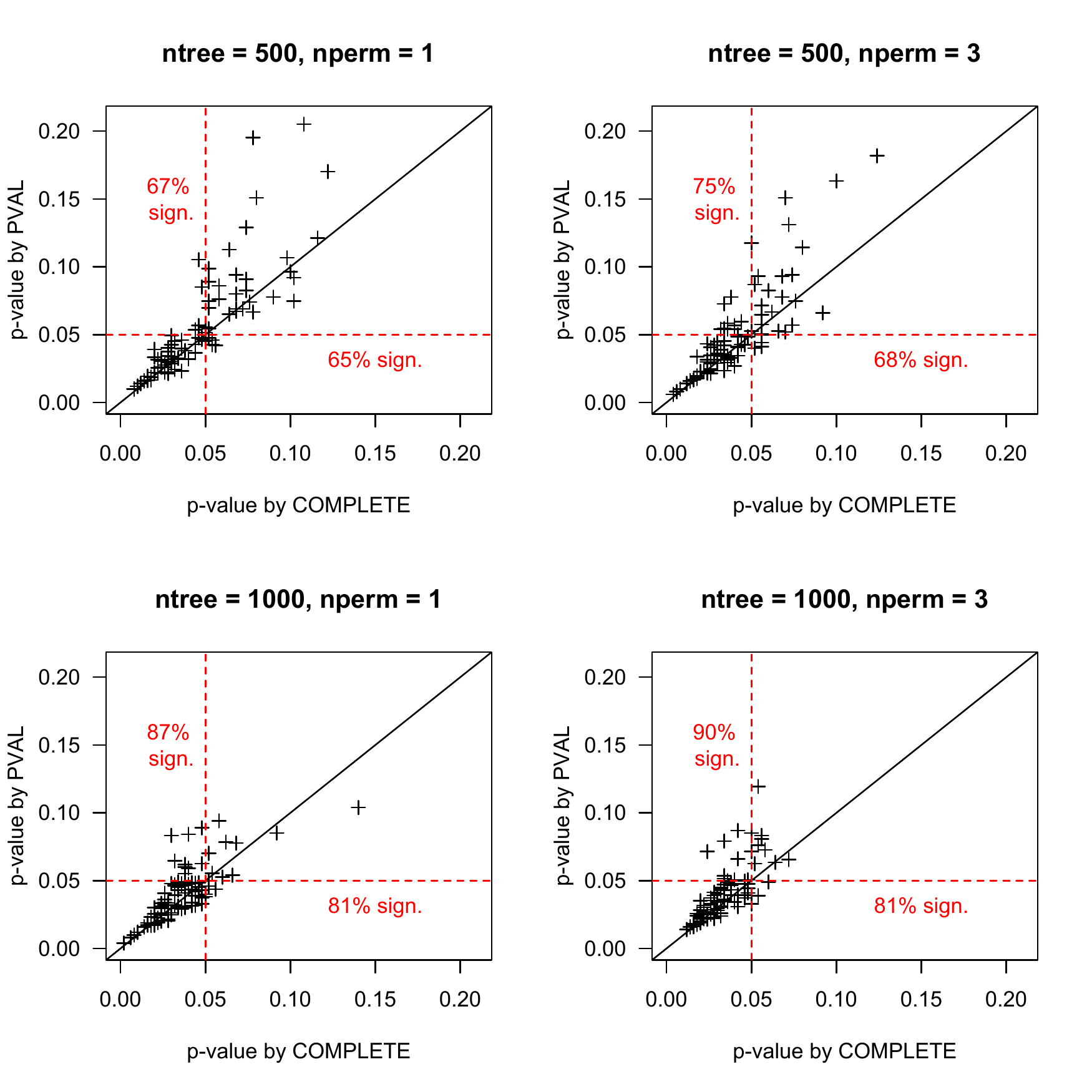}
         \caption{P-values estimated for triceps skin fold thickness by COMPLETE and PVAL ($M=500$, $ntree \in \{500, 1000\}$, $nperm \in \{1, 3\}$).}
         \label{fig:Pima3}
     \end{subfigure}
        \caption{Consistency of the methods' decisions to accept one of the two hypotheses and variability of the estimated p-values in the first application study.}
        \label{fig:Pima}
\end{figure}

Results on the distribution of p-values estimated by COMPLETE with $M=10000$, $nperm=3$ and $ntree=1000$ are displayed in Figure \ref{fig:Pima1}. With 20 applications to the data, consistent decisions to accept either $H_{0}$ or $H_{1}$ can be observed for diastolic blood pressure and diabetes pedigree function, respectively. Despite the high number of permutations and the attempt to generate a stable result over many trees in the RF model and by calculating an average VIMP, the p-values for diastolic blood pressure still scatter considerably. In the case of triceps skin fold thickness, this even leads to divergent decisions with two acceptances of $H_0$ and else only acceptances of $H_1$. The variability of results is therefore further explored in the following for triceps skin fold thickness. In this respect, it is interesting to note that the p-values of the variables scatter differently, since the variance of the Bernoulli distribution is $p(1-p)$. The latter takes its maximum for $p=0.5$ and has the positive property in the given context of tending towards zero for smaller (and larger) values of p.

A comparison of p-values estimated by COMPLETE and PVAL with $M=500$ permutations is given in Figure \ref{fig:Pima3} and shows that COMPLETE more consistently accepts $H_1$ for triceps skin fold thickness. Thereby, both the higher value of $nperm$, but especially the higher value of $ntree$, lead to an increased consistency of the decisions of both methods, which is ultimately alike that of COMPLETE with $M=10000$. A similar pattern emerges for each method with $M=500$ permutations as shown in Figure \ref{fig:Pima2}. The methods more consistently reject $H_{0}$ or accept $H_{1}$ for skin fold thickness when $nperm$ and especially $ntree$ are increased. The most consistent decisions are made by COMPLETE and CERTAIN, followed by SAPT, SPRT and PVAL. The average number of permutations to premature termination also increases accordingly for each method, with PVAL requiring the fewest permutations, followed by SAPT, SPRT and CERTAIN. A direct comparison of SAPT and SPRT again shows that SAPT tends to accept $H_{1}$ more often and earlier for this rather important variable.

\subsection{SARS-CoV-2 Data}

A second application study is based on data from a prospective diagnostic study conducted to investigate the usefulness of self-reported contact history and clinical symptoms in ruling out or detecting SARS-CoV-2 infections in general practice \citep[]{Schneider2021}. Participants were recruited from November 2020 to February 2021 in family practices in urban and rural areas of Bavaria, Germany. The presence or absence of infection poses a classification problem and was established by PCR testing of nasopharyngeal swabs. There are 18 potential diagnostic factors: contact with infected person, stay in Corona risk area, contact with persons with suspected infection, sex, age, anosmia/ageusia, fever, sudden disease onset, limp pain, dry cough, headache, common cold, fatigue, diarrhea, sore throat, dyspnea, nicotine use and chronic disease. All these variables are binary (yes/no), except age, which was measured in years. The original study focused on the accuracy of diagnostic models, including a RF model, and the identification of prognostic factors. The latter is to be extended here through the application of corresponding hypothesis tests regarding the VIMP of the variables. 

In order to establish comparability with the original study, a complete-case analysis is carried out. The data set used comprises 1141 observations. SAPT with $(p_{0} = 0.06,\ p_{1} = 0.04,\ A = 0.1,\ B = 10)^{\top}$ is used in this specific application example to determine diagnostic factors with a statistically significant VIMP. Following the general recommendation for classification problems, the square root of the total number of predictor variables was set to $mtry = 5$. To produce a more stable result, $ntree=1000$ trees were generated in the RF model and an estimate of VIMP averaged over $nperm=3$ permutations was calculated.

\begin{figure}
\centering
\includegraphics[width=0.8\textwidth]{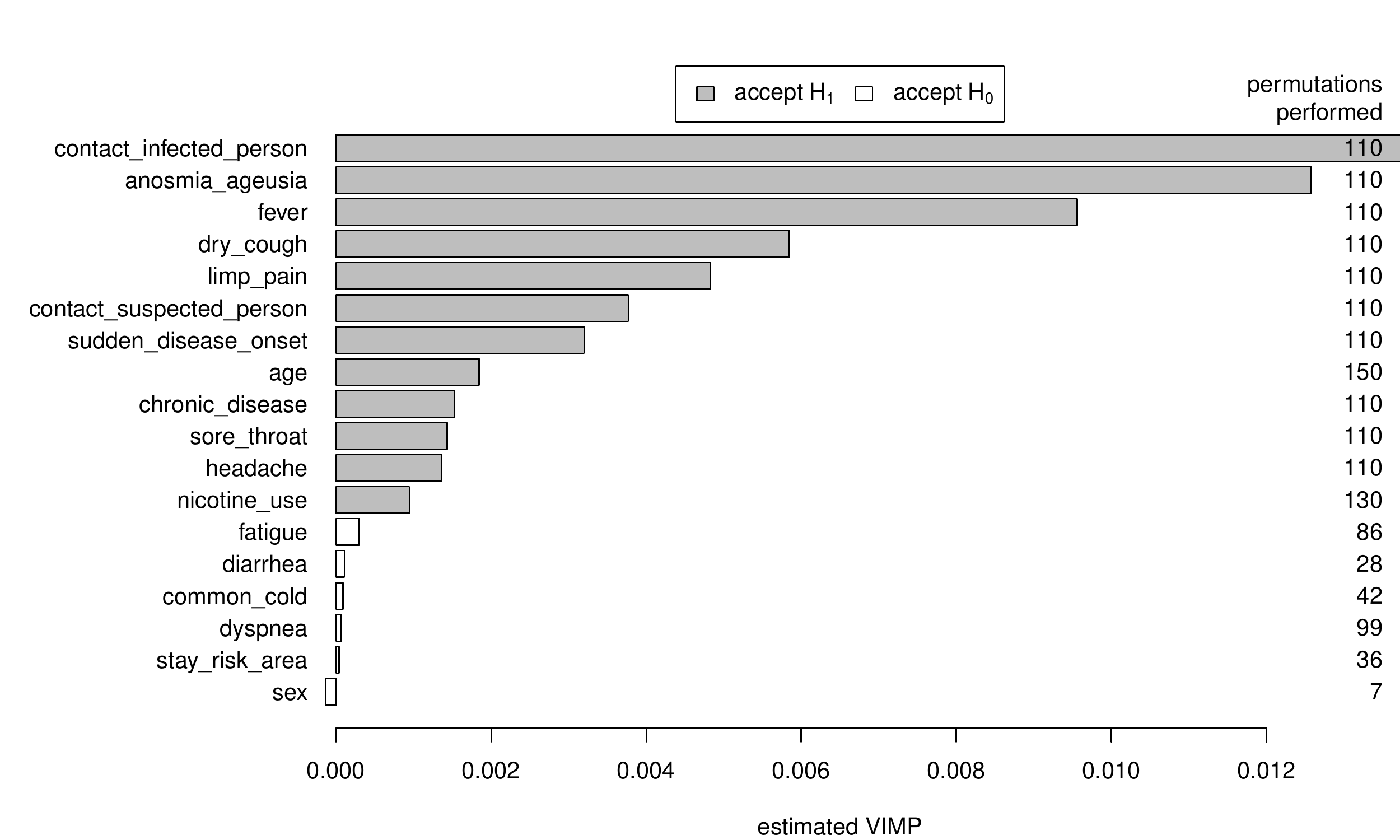}
\caption{Estimated VIMP of the potential prognostic factors of SARS-CoV-2 infection in the second application study. Statistically significant results in which SAPT made the test decision "accept $H_{1}$" are depicted by filled bars. The VIMP of the variable 'contact with infected person' has been truncated for better visibility of all results and is 0.031.}
\label{fig:COVID}
\end{figure}

The results are shown in Figure \ref{fig:COVID}. SAPT terminated prematurely for each variable, accepting $H_{1}$ for twelve variables and $H_{0}$ for six variables. It stopped at $m=110$ permutations for most variables with a high estimated VIMP, which is the earliest possibility to accept $H_{1}$ (cf. Equation \ref{eq:acceptsuccesses}, Equation \ref{eq:rejectsuccesses} and Figure \ref{fig:Bounds}). Similarly, $H_{0}$ can be accepted for the first time at $m=6$ and thus only $m=7$ permutations were used for the variable 'sex'. Interestingly, for the variable 'age' 150 permutations were required to accept $H_{1}$ although it has a higher estimated VIMP than other variables for which only 110 permutations were required. This suggests that not only the strength of an effect plays a role in premature termination, but also some sources of variability, such as the variance of effect estimation and the randomness of permutation testing. In summary, only $m=7$ to $m=150$ permutations of the maximum possible $m=500$ permutations were required. This corresponds to a saving of 98.6\% to 70.0\% in the permutations performed per variable. A total of 1678 (18.6\%) out of a maximum of 9000 permutations were performed overall. The use of SAPT instead of COMPLETE thus reduced the computing time by a factor of about five in the present application example.

\section{Discussion}
\label{sec:disc}

The distribution of an estimated VIMP under $H_{0}$ is difficult to assess. Therefore, it has been suggested to use empirical distribution functions or to assume the (asymptotic) normality of VIMP under $H_{0}$ based on empirical evidence, and to use sampling methods such as bootstrapping and jackknife estimation to obtain an estimate of variance \citep[]{Janitza2018, Ishwaran2019}. Apart from that, a sophisticated formal test of variable importance under regularity conditions has been derived analytically \citep[]{Mentch2016, Mentch2017}. However, these approaches have been found to be computationally expensive or even practically unfeasible. The same applies to permutation tests, which do not require distributional assumptions and can generically be applied to any type of RF and VIMP, but are also very computationally expensive \citep[]{Hapfelmeier2013, Hapfelmeier2014, Nembrini2018, Speiser2019}. An exception to these high computational costs has recently been introduced with a two-sample permutation test, as it only requires the computation of two RF models \citep[]{Coleman2019, McAlexander2020}.

In any case, it is obvious that permutation tests can be terminated prematurely if the decision for $H_{0}$ or $H_{1}$ has already been established before all permutations have been carried out. Thus, corresponding computationally expensive calculations can be skipped. The inequalities (\ref{eq:earlystopp1}) and (\ref{eq:earlystopp2}) in Section \ref{sec:permtest} provide according decision rules and a respective method has been named CERTAIN here. Beyond that, the more sophisticated sequential permutation tests SPRT and SAPT have been proposed to minimize the number of permutations required before a hypothesis can be accepted prematurely while controlling the type-I error probability and controlling or minimizing the type-II error probability \citep[]{Wald1945, Lock1991}. Respective decision rules are given by the inequalities (\ref{eq:acceptsuccesses}) and (\ref{eq:rejectsuccesses}). Another sequential approach, called PVAL here, allows the computation of a maximum likelihood estimate of the p-value of a permutation test using the case dependent equations (\ref{eq:pval}) \citep[]{Besag1991}. The estimated type-I error probability, the type-II error probability (= 1 - Power), the average number of permutations until premature termination and the stability of decisions for $H_{0}$ or $H_{1}$ were examined and compared between these methods in the present work. A permutation test without implementation of an early stopping rule, called COMPLETE here, served as a benchmark. 

The results of two simulation studies show that the proposed heuristic approach, which implements the abovementioned methods through two-sample permutation tests (denoted as *\_TS), leads to a considerably increased type-I error probability. Of course, this goes hand in hand with a high power. Without strict error control, however, the heuristic approach cannot be recommended for hypothesis testing. Nevertheless, it has the advantage of an enormously reduced computing time, since only two RF models have to be generated. Against this background, it can be useful for certain purposes, for example for exploratory analyses, that is for hypothesis generation, or to filter variables in a very liberal approach before further processing. Due to the short computation time, it is also not important to use sequential approaches and the application of COMPLETE\_TS may be feasible in many cases. Due to the lack of error control, the two-sample permutation test is not considered further in the following discussion.

The type-I error probability of the sequential methods SPRT, SAPT and PVAL was well controlled at the nominal significance level. Remarkably, their power was comparable and $\geq 97$\% of COMPLETE's power in the present simulation studies. The observed average number of permutations to premature termination under $H_{0}$, using the chosen specifications of each method, corresponded to the expected and very low values of 35.4, 43.6 and 40.6 for SPRT, SAPT and PVAL, respectively. In general, the relevance of a variable to the prediction problem directly affected the average number of permutations required until each method terminated prematurely. For strong or informative predictor variables, SAPT was superior to SPRT. For weak or uninformative predictor variables, the opposite was true. To be able to stop prematurely, PVAL and CERTAIN depend strongly on observing many statistics $u_{s} \geq u$, that is on observing a high $d_{m}$. This is only the case for weak predictor variables and corresponds to an early acceptance of $H_{0}$, but not of $H_{1}$. Overall, SPRT and SAPT required the lowest average number of permutations.

A general matter of concern is the inconsistency of the methods' decisions to accept $H_{0}$ or $H_{1}$ for a specific variable using a given dataset. It results from three sources of randomness, namely the permutations performed for the permutation test \citep[][]{Lock1991}, the sampling of data and variables used to build the RF model, and the permutation of out-of-bag data for the estimation of VIMP. To solve the latter two issues and consequently obtain a deterministic estimate of VIMP, one would have to use the infinite $VIMP_{\Theta}(X_{j})$ (cf. equation (\ref{eq:vimpinf})) instead of $VIMP(X_{j})$ (cf. equation (\ref{eq:vimp})) for permutation testing. However, $VIMP_{\Theta}(X_{j})$ requires fitting a RF model with $T \to \infty$ trees, which is impossible and can only be approximated by choosing a high number of trees. A more stable estimate of VIMP can also be obtained by averaging multiple estimates over several permutations of the out-of-bag data. Therefore, high values should be chosen for the corresponding parameters $ntree$ and $nperm$ in the RF algorithm. The first application study carried out shows that the consistency of the results can be considerably increased in this way, especially by increasing $ntree$. Inconsistency is less of a problem with very weak or strong predictor variables. In these cases, SPRT and SAPT tend to clearly accept $H_{0}$ or $H_{1}$. An examination of the very steep power functions in Figure \ref{fig:Power} supports this finding, because for $p<p_{1}$ and $p>p_{0}$ the probability of rejecting or accepting $H_{0}$ quickly approaches 1 or 0, respectively. Similarly, the p-values estimated by PVAL follow a Bernoulli distribution with variance p(1 - p) and therefore tend to be more consistent for smaller (and larger) values of p.  

A second application study showed an example of a practical application of SAPT to data which was originally collected for the diagnosis of SARS-CoV-2 infection by self-reported contact history and clinical symptoms \citep[]{Schneider2021}. SAPT terminated prematurely for each variable. Only 18.6\% of permutations compared to COMPLETE with $M=500$ permutations had to be performed to accept $H_{0}$ and $H_{1}$ for six and twelve variables, respectively.

A limitation of the present study is that it focuses on hypothesis tests with a nominal significance level of $\alpha = 0.05$. While this is a common choice, applicants may wish to use other values. Adjusting for the multiple testing problem by controlling for the family-wise error rate (FWER) or false discovery rate (FDR) is a related topic. COMPLETE, CERTAIN and PVAL can be readily applied as described. It is also possible to adapt the specifications of SPRT and SAPT accordingly, but this is beyond the scope of the present work. It is planned that future implementations of the methods will cover these cases. Furthermore, it should be mentioned that there are situations in which the sequential methods may not need to be applied. For example, COMPLETE can be used alternatively if there is sufficient time to wait for the results of the analysis or if the application to few predictor variables or few observations allows a fast calculation. However, the latter is not the usual or recommended application area for data hungry machine learning methods \citep[]{Ploeg2014, Riley2020}. The out-of-bag error estimates commonly used in the calculation of VIMP have been criticised for possible bias, but an empirical study co-authored by the second author suggested that this bias may only be relevant for small samples \citep[][]{Janitza2018b}.

\section{Conclusion}
\label{sec:conc}

Considering that SPRT and SAPT seem to be associated with an error control and power that is almost equally as good as COMPLETE, these two methods can be recommeded for accepting $H_{0}$ or $H_{1}$ prematurely when testing the statistical significance of VIMP. Thereby, SPRT and SAPT require fewer permutations under $H_{0}$ and $H_{1}$, respectively. The applicant should decide which property is more suitable for his or her application. For example, SAPT would be more efficient than SPRT if it can be assumed that a predominant proportion of the variables in the data are relevant to the prediction problem. In the case of few relevant variables, SPRT would be more efficient. PVAL should be used when an estimate of a p-value is needed. However, it has the disadvantage that it stops early only with rather weak predictor variables. CERTAIN leads to the same test decisions as COMPLETE, but yields the smallest savings in the number of permutations. CERTAIN is most efficient for weak predictor variables. For each method, a high number of trees should be chosen in the RF model to obtain consistent decisions for acceptance of $H_{0}$ or $H_{1}$. Although the approach is proposed here for RF, it translates well to any kind of prediction model. A very fast heuristic approach based on a two-sample permutation test can be considered for exploratory purposes, but is associated with an increased type-I error probability.

In conclusion, the proposed sequential p-value estimation and permutation tests of VIMP are theoretically well-founded and less computationally expensive than conventional permutation testing. An implementation is made available via the associated R package $rfvimptest$ on the Comprehensive R Archive Network (CRAN).

\appendix

\section{R Code}
\label{app:AppA}
R Code of the methods and the performed simulation and application studies is provided as supplementary material.

The associated R package $rfvimptest$ is available from CRAN.

\section{Acknowledgements}
We would like to thank Anne-Laure Boulesteix for very useful feedback and discussions.

\bibliographystyle{unsrt}  
\bibliography{Mybiblio.bib}  

\end{document}